\def\papertitle{Hyper Recurrent Neural Network: Condition Mechanisms for Black-box Audio Effect Modeling}
\def\paperauthorA{Yen-Tung Yeh}
\def\paperauthorB{Wen-Yi Hsiao}
\def\paperauthorC{Yi-Hsuan Yang}
\documentclass[twoside,a4paper]{article}
\usepackage{etoolbox}
\usepackage{dafx_24}
\usepackage{amsmath,amssymb,amsfonts,amsthm}
\usepackage{euscript}
\usepackage[T1]{fontenc}
\usepackage[utf8]{inputenc}
\usepackage{ifpdf}
\usepackage[english]{babel}
\usepackage{caption}
\usepackage{subfig} % or can use subcaption package
\usepackage{color}
\usepackage{booktabs}
\usepackage{multirow}
\usepackage{soul}

\input glyphtounicode
\ninept

\newcounter{numauth}
\newcounter{listcnt}
\newcommand\authcnt[1]{\ifdefined#1 \stepcounter{numauth} \fi}

\newcommand\addauth[1]{
\ifdefined#1 
\stepcounter{listcnt}
\ifnum \value{listcnt}<\value{numauth}
\appto\authorslist{, #1}
\else
\appto\authorslist{~and~#1}
\fi
\fi}
\authcnt{\paperauthorB}
\authcnt{\paperauthorC}
\authcnt{\paperauthorD}
\authcnt{\paperauthorE}
\authcnt{\paperauthorF}
\authcnt{\paperauthorG}
\authcnt{\paperauthorH}
\authcnt{\paperauthorI}
\authcnt{\paperauthorJ}
\def\authorslist{\paperauthorA}
\addauth{\paperauthorB}
\addauth{\paperauthorC}
\addauth{\paperauthorD}
\addauth{\paperauthorE}
\addauth{\paperauthorF}
\addauth{\paperauthorG}
\addauth{\paperauthorH}
\addauth{\paperauthorI}
\addauth{\paperauthorJ}
\usepackage{times}
\newif\ifpdf
\ifx\pdfoutput\relax
\else
   \ifcase\pdfoutput
      \pdffalse
   \else
      \pdftrue
\fi

\ifpdf % compiling with pdflatex
  \usepackage[pdftex,
    pdftitle={\papertitle},
    pdfauthor={\authorslist},
    pdfsubject={Proceedings of the 27th International Conference on Digital Audio Effects (DAFx24)},
    colorlinks=false, % links are activated as color boxes instead of color text
    bookmarksnumbered, % use section numbers with bookmarks
    pdfstartview=XYZ % start with zoom=100% instead of full screen; especially useful if working with a big screen :-)
  ]{hyperref}
  \usepackage[pdftex]{graphicx}
\else % compiling with latex
  \usepackage[dvips]{epsfig,graphicx}
  \usepackage[dvips,
    pdftitle={\papertitle},
    pdfauthor={\authorslist},
    pdfsubject={Proceedings of the 27th International Conference on Digital Audio Effects (DAFx24)},
    colorlinks=false, % no color links
    bookmarksnumbered, % use section numbers with bookmarks
    pdfstartview=XYZ % start with zoom=100% instead of full screen
  ]{hyperref}
\fi
\usepackage[hypcap=true]{caption}
\title{\papertitle}

\threeaffiliations{
\paperauthorA \,\thanks{\vspace{-3mm}}}
{\href{https://affige.github.io/lab.html}{Music and AI Lab} \\ National Taiwan University\\ Taipei, Taiwan\\
{\tt \href{mailto:r12942179@ntu.edu.tw}{r12942179@ntu.edu.tw}}
}
{\paperauthorB \,\thanks{\vspace{-3mm}}}
{Indepedent Researcher \\ Taipei, Taiwan \\ {\tt \href{mailto:s101062219@gmail.com}{s101062219@gmail.com}}
}
{\paperauthorC \,\thanks{\vspace{-3mm}}}
{\href{https://affige.github.io/lab.html}{Music and AI Lab} \\ National Taiwan University\\ Taipei, Taiwan\\
{\tt \href{mailto:yhyangtw@ntu.edu.tw}{yhyangtw@ntu.edu.tw}}
}

\begin{document}
% more pdf-tex settings:
\ifpdf % used graphic file format for pdflatex
  \DeclareGraphicsExtensions{.png,.jpg,.pdf}
\else  % used graphic file format for latex
  \DeclareGraphicsExtensions{.eps}
\fi

%\makeatletter
%\pdfbookmark[0]{\@pdftitle}{title}
%\makeatother
\newcommand{\ytrevised}[1]{{\color{blue}#1}}

\maketitle

\begin{abstract}

Recurrent neural networks (RNNs) have demonstrated impressive results for virtual analog modeling of audio effects. These networks process time-domain audio signals using a series of matrix multiplication and nonlinear activation functions to emulate the behavior of the target device accurately. To additionally model the effect of the knobs for an RNN-based model, existing approaches integrate control parameters by concatenating them channel-wisely with some intermediate representation of the input signal. While this method is parameter-efficient, there is room to further improve the quality of generated audio because the concatenation-based conditioning method has limited capacity in modulating signals. In this paper, we propose three novel conditioning mechanisms for RNNs, tailored for black-box virtual analog modeling. These advanced conditioning mechanisms modulate the model based on control parameters, yielding superior results to existing RNN- and CNN-based architectures across various evaluation metrics. 
\end{abstract}

\section{Introduction}

% What is audio effect modeling
% categorized as white, greedy, black-box approach: brief overview 
% common architectures used as black-box 
Audio effect modeling \cite{5280324,imort22ismir} involves creating algorithms or models that replicate the behavior of specific audio effects to emulate vintage hardware \cite{VolterraSeries10, VolterraSeries18, HMModel} or digital audio effect chains \cite{steinmetz2020automatic}. This technique is called the digital emulation of audio effects, also known as virtual analog (VA) modeling. Methods for VA modeling can be categorized into white-box, grey-box, and black-box approaches. White-box methods \cite{diffWhiteBox, Parker:2022.PhysicalModeling} typically require complete knowledge of the target system, achieving high-quality emulation but requiring a time-consuming design process. Grey-box approaches \cite{eichas2018jaes, wright2022grey, colonel2022reverse, miklanek2023neural, nercessian2021lightweight, 3e3037cffa094d3fb7a73dbc33ea0e9a} introduce inductive bias of the system using input-output measurements, allowing flexibility while maintaining interpretability. However, understanding the target device remains crucial and may not be always attainable. To get rid of reliance on prior knowledge constraints, black-box approaches have recently gained popularity for efficient VA modeling, relying solely on device measurements. Black-box approaches often use neural networks to model the target device. In the literature, this active research field proposes mainly three architectures: convolutional-based (CNN) \cite{app10020638, Damskgg2019RealTimeMO,8682805,8683529,steinmetz2022efficient,https://doi.org/10.48550/arxiv.2211.00497}, recurrent-based (RNN) \cite{6567472,schmitz2018real,Zhang2018AVG,769f627fa4fe49569bd207f6b1d32dc3} neural networks, and Neural Ordinary Differential Equations (Neural ODEs) \cite{NeuralODE}. 

% Why conditioning is necessary 
% Previous conditioning methods for CNN and RNN 
% no conditioning methods in Neural ODE 
To accurately and fully replicate the behavior of devices, it is essential to consider the control parameters, a.k.a. knob values, in audio effect emulation. Neural networks typically represent control knob values by conditioning vectors, injecting conditioning information via a certain conditioning mechanism. For CNN-based architectures, different conditioning mechanisms have been studied, such as 
local conditioning \cite{8682805,rethage2018wavenet} and feature-wise linear modulation (FiLM) \cite{perez2017film}.
%. For example, Damskägg \emph{et al.} \cite{8682805} employ the local conditioning strategy proposed in \cite{rethage2018wavenet}, while Steinmetz \emph{et al.} \cite{steinmetz2022efficient} utilizes the feature-wise linear modulation (FiLM) \cite{perez2017film} conditioning module. 
For RNN-based architectures, however, the prevailing conditioning method studied in the literature remains to be the simple concatenation-based method, which simply concatenates the conditioning vector channel-wisely with some intermediate representation of the input audio. For Neural ODEs-based architectures, the employment of conditioning mechanisms has not been studied, to the best of our knowledge \cite{NeuralODE}.

% why not concat? 
% brief overview the core idea of conditioning: modulation 

% Although the concatenation method is parameter-efficient, it limits the modulation potential of the neural network, interrupting the complete optimization of conditioning information. Motivated by exploring the full modulation potential of recurrent neural networks, we aim to design advanced conditioning mechanisms. 
Strengths of the concatenation method include its parameter efficiency, simplicity, and ease of implementation. However, it has the downside of being too simple to provide enough capacity to model complicated input/output relationships, resulting in limited modeling performance. Taking inspiration from the work of Richard \emph{et al}. \cite{2021hyeprconv}, who use the hypernetwork \cite{DBLP:journals/corr/HaDL16} to use the conditioning information to generate conv1d weights for their CNN-based model for mono-to-binaural synthesis, we aim to explore the application of hypernetworks to harness control parameters to adapt the weights of RNN models for audio effect modeling. This adaptation can be achieved through either generation or modulation by the output of another neural network. Thus, we investigate using hypernetwork variants as conditioning mechanisms for virtual analog modeling of audio effects. 

% transient 
Besides, we note that previous research \cite{simionato2022deep} has presented various examples indicating that an RNN model may have limitations in modeling the ``transients'' for compressor modeling. Motivated by this observation, we proposed a metric to objectively evaluate transient reconstruction loss based on the transient modeling synthesis method (TMS) \cite{verma2000extending}. These objective results offer insight into a model's complete transient modeling capability.

% contribution 
Accordingly, our work presents three main contributions: firstly, we propose three hypernetwork-based conditioning methods for RNNs to handle control parameters. We demonstrate that all proposed conditioning methods outperform the concatenation method through objective evaluation and show lower training compute. Secondly, we introduce a new objective evaluation metric for estimating transient reconstruction error. Finally, we compared CNN-based and RNN-based models with different conditioning methods. The results show that the proposed method for RNNs can achieve better audio quality and more accurate transient reconstruction. We provide audio samples online,\footnote{\url{https://yytung.notion.site/HyperRNN}} and share the source code with an open-source license.

% https://yytung.notion.site/HyperRNN-ce8cecb3ef6a41fcbcb54a94f9f6de6f
% https://bit.ly/3TjASFe
% structured
%The paper is structured as follows: Section 2 reviews various black-box audio effect modeling methods, while Section 3 introduces the target device we aim to emulate and outlines the utilization of training data. Section 4 presents the proposed conditioning modules, and Section 5 explains the experimental setup. Objective evaluation results are reported in Section 6, with further discussion and future work in Section 7. \ytrevised{As a complement of this paper, we provide audio samples online \footnote{\url{https://bit.ly/3TjASFe}}, and we will open source our code upon paper publication.}

\section{Methods for black-box modeling}

% consideration of the model architecture: performance, real-time usage
Black-box modeling approaches can be achieved using different architectures, e.g., CNNs, RNNs, and neural ODEs. Each architecture has its advantages and disadvantages with mainly two considerations: model performance and real-time usage. 
%Three architectures are proposed in prior studies: CNNs, RNNs, and neural ODEs. 

% introduce CNN-based 
Many CNN-based models for VA modeling
%have been proposed. Many of them 
are modified from WaveNet \cite{oord2016wavenet}, the famous architecture for processing time-domain signals. The model's advantages are the high quality of emulation, parallel computation, and fast inference time running on GPUs. However, when considering real-time usage on CPUs,  CNNs tend to be slower than  RNNs \cite{769f627fa4fe49569bd207f6b1d32dc3}. Another concern is the high latency. As mentioned in \cite{simionato2023fully}, the lower bound of the latency of CNNs is the size of the receptive field. When the target effect requires a large receptive field to achieve better quality, such as compressor \cite{https://doi.org/10.48550/arxiv.2211.00497}, the high latency problem will harm real-time usage.

% introduce RNN-based 
%Current 
RNN-based models are usually based on long-short term memory (LSTM) or gated recurrent units (GRU). Owing to their recurrent nature, both architectures can have access to information from the past and accordingly excel in modeling sequential data. These networks demonstrate high quality in VA modeling while requiring fewer parameters compared to CNNs \cite{769f627fa4fe49569bd207f6b1d32dc3, steinmetz2022efficient}. Additionally, they boast fast inference times and low latency for real-time applications because of their step-by-step mechanism, which aligns with the real-time audio input fed to the system sample-by-sample. Despite their real-time performance advantages, RNN-based models encounter several challenges. First, unstable training is a common issue attributed to vanishing or exploding gradients, leading to increased development efforts in model design. Second, unlike CNN-based models, RNN-based models cannot leverage parallel computation due to their recurrent behavior, thus missing out the benefits of GPU acceleration, leading to longer training time.  

% introduce Neural ODE
Neural ODEs use the ODE mechanism to emulate the first- and second-order diode-clipper \cite{NeuralODE}. Neural ODE can achieve performance comparable to RNN-based neural networks but with fewer parameters. Due to its properties, Neural ODE can achieve arbitrary sample rates, which indicates that it can save the computation effort of resampling, which the previous architecture cannot. While the method shows promising results, it has not been tested on complex systems such as the pedal or the amp. 

%\section{Modelled devices}

\section{Dataset}

We consider two datasets and the modeling of two types of effects in this study. We provide some details below.

%VA modeling is widely employed to simulate nonlinear audio circuits, including distortion pedals, guitar amplifiers, and analog compressors. 

% why two devices -> different types of audio effects -> proposed methods generalizability 
%VA modeling is commonly employed to simulate nonlinear audio circuits, including distortion pedals, guitar amplifiers, and analog compressors. In this study, we chose to model two devices: the Boss OD-3 overdrive pedal and the Teletronix LA-2A compressor. According to the taxonomy of audio effects presented in \cite{martinez2020deep}, we can consider the Boss OD-3 overdrive pedal as a type of \emph{nonlinearity with short-term memory}, while the LA2A compressor is viewed as a \emph{nonlinearity with long-range dependencies}. Employing these two distinct device types enables us to assess the strengths and weaknesses of the proposed approach more effectively. 

\subsection{Teletronix LA-2A compressor}
\label{sec:db:db1}

Existing datasets \cite{pedroza2022egfxset, stein2010automatic} typically provide specific device settings (a.k.a., ``snaptshots''). However, our task requires additionally a range of control parameter information. Hence, we chose the Teletronix LA-2A compressor as a target device. The Teletronix LA-2A compressor has been widely used in previous studies on VA  modeling of compressors \cite{steinmetz2022efficient, wright2022grey}, and the dataset was compiled by Hawley \emph{et al.} \cite{hawley2019signaltrain}. As outlined in the audio effects taxonomy presented in \cite{martinez2020deep}, the LA-2A compressor is categorized under \emph{nonlinearity with long-range dependencies}.

The behavior of the LA-2A compressor is governed by two primary parameters: the switching control and the peak reduction control. The switching control determines whether the LA-2A operates in limit or compress mode. Meanwhile, the peak reduction control knob controls the degree of compression applied to the signal. 
%It is noteworthy that the LA-2A lacks attack and release controls. 
Their input signal included noise and various instrument clips, ensuring comprehensive coverage of the device's behavior. The dataset consists of approximately 20 hours of recordings at a sampling rate of 44.1kHz. In our research, we utilized a specific subset of this dataset, concentrating exclusively on the compress mode of the data. This subset encompasses peak reduction values ranging from 0 to 100 in increments of 10, following the settings outlined in \cite{wright2022grey}. We partitioned the dataset using an 80/10/10 ratio for the train/validation/test sets. Each conditioning information is encountered during training while varying audio contents test the model's generalizability at the inference stage. 

\subsection{Boss OD-3}
\label{sec:db:db2}

It is vital to assess our methods across different effect types. Because we already have the LA-2A compressor, which is a type of effect with long-range dependencies, we aim to include a device of \emph{nonlinearity with short-term memory} types, such as the overdrive pedal. To our knowledge, there is no publicly available fully-conditioned overdrive pedal dataset. Hence, given its status as a classic overdrive pedal, we gathered data from the Boss OD-3 overdrive pedal on our own. 
%We provide more details below.

The Boss OD-3 pedal is a famous overdrive pedal, initially introduced in 1997.\footnote{\url{https://www.boss.info/us/promos/40th_anniversary_compact_pedals/}} A subsequent pedal version was released in 2021, featuring only minor differences. For this study, we focus on modeling the 1997 pedal version. %Boss OD-3 is based on soft-clipping circuits, so its response will be smoother and more natural than hard-clipping-based circuits. The pedal is well known for its broader range of timbre because it can respond to player dynamics sensitively. 
It is equipped with three distinct knobs, offering precise control over its operational parameters. The ``level'' knob regulates post-gain, determining the output volume after the nonlinear clipping stage. The ``tone'' knob adjusts equalization by blending bass and treble frequencies, influencing perceived brightness or darkness. Finally, the ``gain'' knob determines the degree of distortion applied to the signal, acting as a pre-gain mechanism amplifying the input signal before clipping takes place. We collected the conditioned Boss OD-3 dataset on our own with the following specifics. Among the three control parameters of Boss OD-3, we do not consider the ``level'' control, for such an effect can be readily achieved in the digital domain by multiplying with a constant. For ``tone'' and ``gain'', we segmented their control range into five equal intervals, providing each knob with five distinct control values, from 0 to 4, each representing the index of the interval. Recordings were directly from the Boss OD-3 device, using signals such as white noise, guitar, bass, drum loops, and vocals as the input signals. Each input signal was about 6 minutes long and was recorded at a 48kHz sampling rate. In total, our dataset comprises approximately 150 minutes and includes 25 cases, with each tone and gain knob offering five different control values. We divided the dataset into training, validation, and test sets using an 80/15/5 ratio. The model has been trained using all conditioning information during training, while varying audio contents test the model's generalizability at the inference stage. 
%However, the audio content in the test and validation sets is not encountered during training. This splitting strategy evaluates a model's generalizability, probing how well it learns the conditioning information. 

%and to further explore fully-conditioned VA modeling, 
We share this dataset publicly for reproducibility.
Please visit the demo page to find the link.

\section{Proposed approach}

%The proposed conditioning mechanisms are based on hypernetworks \cite{DBLP:journals/corr/HaDL16}. The core idea of hypernetworks is to generate neural network (target network) weights by another neural network (hypernetwork). The original purpose of the hypernetwork is not for the conditioning, but in our task, we apply it as the conditioning mechanism. 

While the proposed conditioning methods can be in general applied to most RNN architectures, for simplicity we consider the case of using the standard RNN below to introduce our methods. The standard RNN formulation is as follows:
\[h_{t} = \tanh(W_{h}h_{t-1} + W_{x}x_{t} + b)  \]
Here, $h_{t}$ represents the hidden state at the $t^{th}$ step and also the output from the current step of the RNN cell. The two weight matrices, $W_{h}$  and $W_{x}$, are for the previous hidden state $h_{t-1}$ and the input signal $x_{t}$, respectively. Additionally, there is one bias vector $b$. The feature maps calculated from $W_{h}$ and $W_{x}$ are denoted as $\mathbf{F}^{h}_{c}$ and $\mathbf{F}^{x}_{c}$. 
\begin{alignat*}{2}
    \mathbf{F}^{h}_{c}=W_{h}h_{t-1}, \quad && \mathbf{F}^{x}_{c}=W_{x}x_{t}
\end{alignat*}
We note that the weight and bias remain fixed throughout the entire time sequence, a concept known as weight-sharing \cite{Sherstinsky_2020}.

\subsection{FiLM-RNN}

%Feature-wise Linear Modulation (FiLM), as denoted in \cite{perez2017film}, is the form of the hypernetwork. 

The first method uses the feature-wise linear modulation (FiLM). While FiLM has been used as the conditioning module for CNN-based models such as Mirco-TCN \cite{steinmetz2022efficient}, it has not been applied to the RNN-based models for VA modeling, to our best knowledge. The FiLM layer's objective is to modulate the target network based on the conditioning input signal. Specifically, FiLM involves two steps: the FiLM-ed generator and FiLM-ed operation. Given the conditioning signal $\phi$, the FiLM-ed generator aims to learn two functions, $f$ and $g$, which output the coefficients $\alpha_{i, c}$ and $\beta_{i, c}$:
\begin{alignat*}{2}
    \alpha_{i, c}=f(\phi), \quad && \beta_{i, c}=g(\phi)
\end{alignat*}
$\alpha_{i, c}$ and $\beta_{i, c}$ are then applied to modulate the feature map $\mathbf{F}_{i, c}$ via feature-wise linear transformation, termed the FiLM-ed operation:
\[FiLM(\mathbf{F}_{i, c}, \alpha_{i, c}, \beta_{i, c}) = \alpha_{i, c}  \mathbf{F}_{i, c} + \beta_{i, c}\] 
The subscripts in $\alpha_{i, c}$, $\beta_{i, c}$, and $\mathbf{F}_{i, c}$ refer to the $c^{th}$ input feature for the $i^{th}$ layer. We discarded the subscript $i$ because the architecture used for VA modeling often uses only one single layer of the recurrent cell \cite{769f627fa4fe49569bd207f6b1d32dc3}. In practice, functions $f$ and $g$ are achieved by few neural layers and are learned end-to-end from the data. 

\begin{figure}[t] 
\centering
\includegraphics[width=0.7\columnwidth]{ 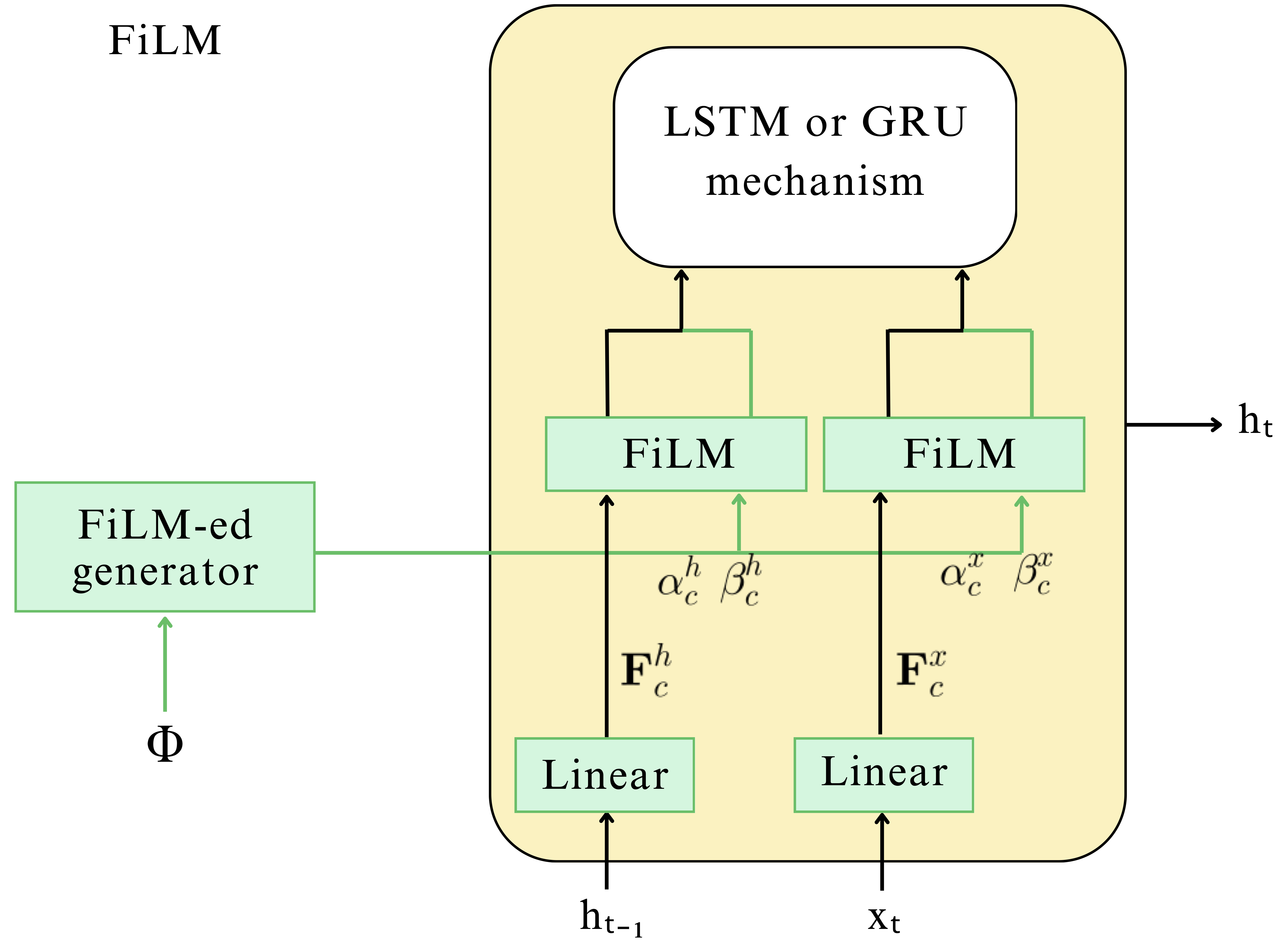} 
\caption{The architecture of the FiLM-RNN, with $\phi$ representing the conditioning vector, $h$ representing the hidden state, $x$ denoting the input signa. The FiLM-ed generator aims to produce scaling and shifting coefficients for feature-wise linear modulation of the feature maps.
}
\label{fig:film}
\end{figure}

As depicted in Figure \ref{fig:film}, we inject the external conditioning vector $\phi$ into the FiLM-ed generator. The FiLM-ed generator will predict two groups of scaling coefficients and shifting coefficients, ($\alpha^{h}_{c}$, $\beta^{h}_{c}$), ($\alpha^{x}_{c}$, $\beta^{x}_{c}$), corresponding to $\mathbf{F}^{h}_{c}$ and $\mathbf{F}^{x}_{c}$. These coefficients will be computed with the FiLM-ed operation to modulate the model's behavior with the corresponding control parameters. 
\begin{align*}
    FiLM(\mathbf{F}^{h}_{c}, \alpha^{h}_{c}, \beta^{h}_{c}) = \alpha^{h}_{c}  \mathbf{F}^{h}_{c} + \beta^{h}_{c} \\
    FiLM(\mathbf{F}^{x}_{c}, \alpha^{x}_{c}, \beta^{x}_{c}) = \alpha^{x}_{c}  \mathbf{F}^{x}_{c} + \beta^{x}_{c}
\end{align*}

%%%%%%%%

\subsection{StaticHyper-RNN}\label{subsec:static}

\begin{figure}[t] 
\centering
\includegraphics[width=0.7\columnwidth]{ 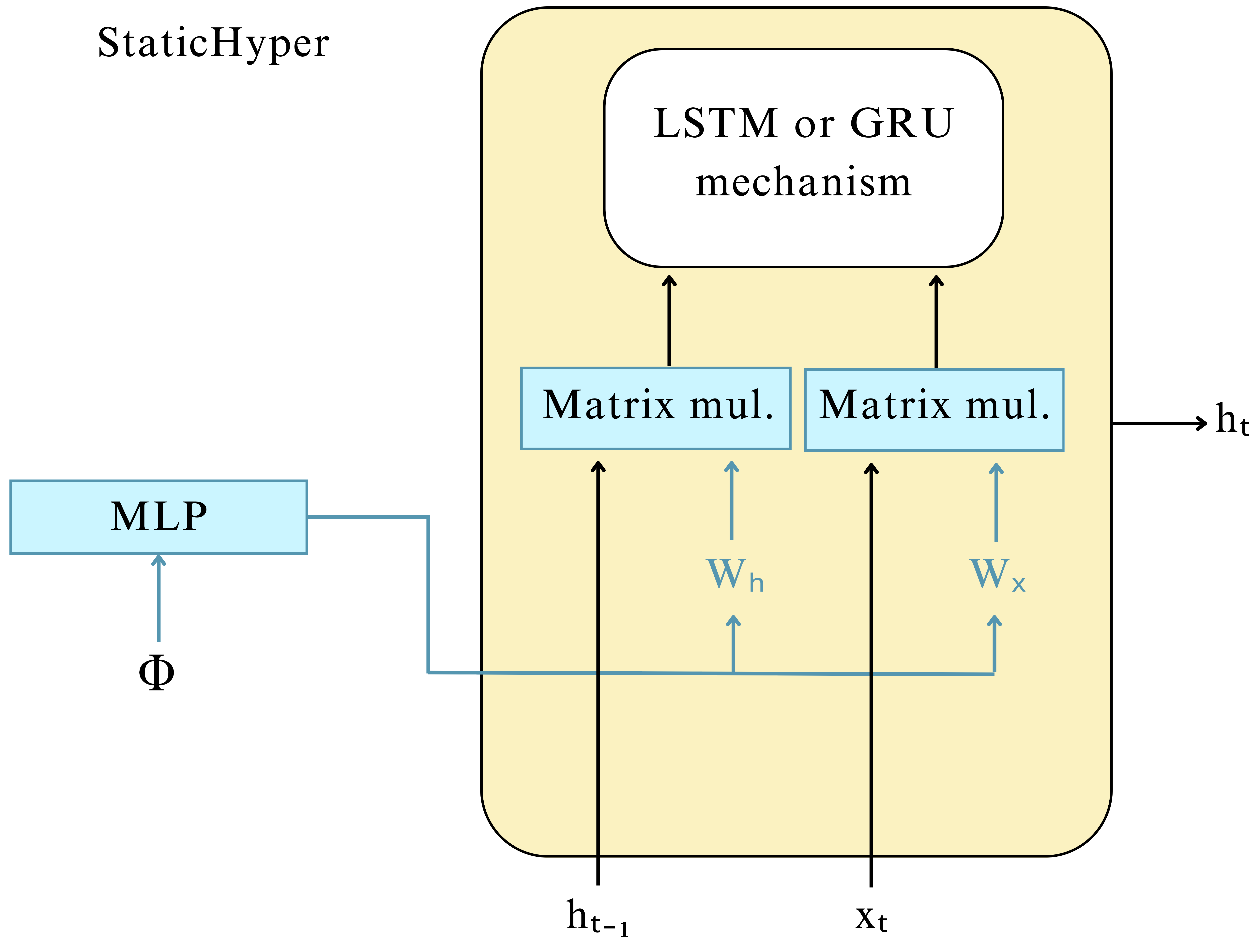} 
\caption{The architecture of the StaticHyper-RNN, with $\phi$ representing the conditioning vector, $h$ representing the hidden state, $x$ denoting the input signal. The MLP aims to generate the weight matrix $W_h$ and $W_x$ to perform matrix multiplication. 
}
\label{fig:static}
\end{figure}

%%%%%%%%%%%%%%%%%%%%%%
%\ytrevised{To adjust the weights of RNNs, we generate weights based on control parameters. This process resembles the concept of a hypernetwork, although the original purpose of a hypernetwork is not for conditioning \cite{DBLP:journals/corr/HaDL16}. In our study, we employ the hypernetwork to determine the weights of RNNs based on the provided conditioning value.} 

FiLM uses the conditional signals to modulate the feature maps $\mathbf{F}$. In other words, the conditional signals do not affect the weight matrices $W$. In contrast, the idea of hypernetwork is to affect the weight matrices $W$ directly. Depending on the conditional signals, the RNN would use different weight matrices $W$ to process an input signal, as shown in Figure \ref{fig:static}. Specifically, the proposed conditioning mechanism is detailed as follows: given a conditioning vector $\phi$, the mechanism aims to learn the functions $f_{x}$, $f_{h}$, and $f_{b}$ to generate the weight matrices $W_{x}$, $W_{h}$, and the bias vector $b$.
% \begin{alignat*}{2}
% f_{x}(\phi) &= W_{x}, \quad && f_{h}(\phi) = W_{h} \quad && f_{b}(\phi) = b
% \end{alignat*}
\begin{align*}
f_{x}(\phi) &= W_{x}, &
f_{h}(\phi) &= W_{h}, &
f_{b}(\phi) &= b
\end{align*}
The target network, which takes the input signal and hidden state as the input, only provides the matrix operation without learning the weight matrix itself, and the functions $f_{x}$, $f_{h}$, and $f_{b}$ are learned to generate the target matrix through stochastic gradient descent. These functions are typically implemented as a neural network, e.g., using a multi-layer perceptron (MLP) architecture. 

The key differences among standard RNN, LSTM, and GRU are the size of the learnable weights and the additional mechanisms. For LSTM, the hypernetwork generates four weights corresponding to the four gates of LSTM, while for GRU, the model generates three weights. The mechanism is called StaticHyper-RNN because the proposed approach generates weights once and maintains them fixed across the entire sequence.

\subsection{DynamicHyper-RNN}\label{dynamicchyper}

\begin{figure}[ht] 
\centering
\includegraphics[width=0.7\columnwidth]{ 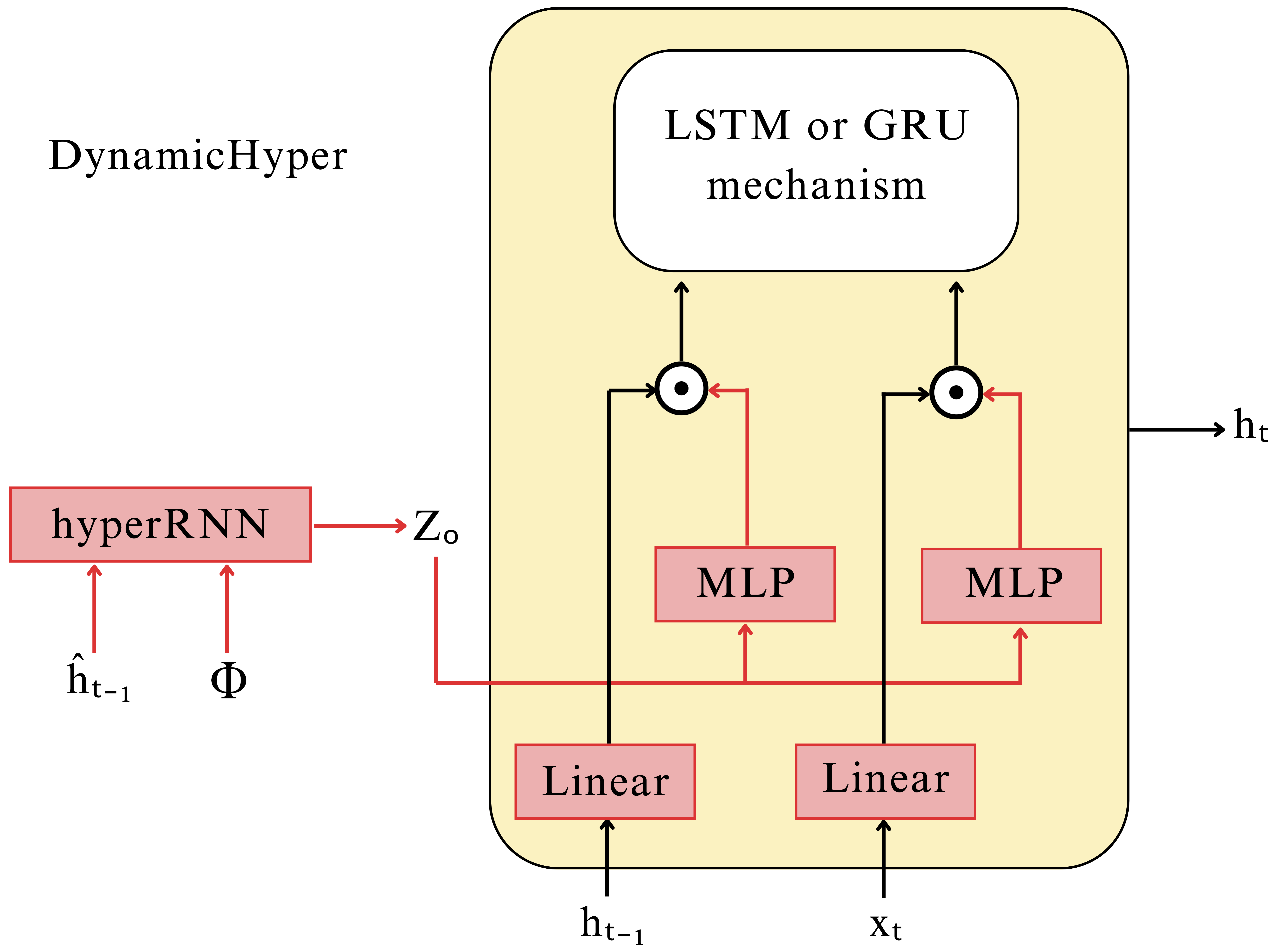} 
\caption{The architecture of the DynamicHyper-RNN mechanism: $\phi$ representing the conditioning vector, $h$ representing the hidden state of the \textbf{mainRNN}, $x$ denoting the input signal, and $\hat{h}$ representing the hidden state of the \textbf{hyperRNN}. The hyperRNN generates the feature $Z_{o}$, then learns an additional transformation to modulate the output of the feature map generated from the input $h$ and $x$.
}
\label{fig:dynamic}
\end{figure}

The concept of a ``dynamic'' hypernetwork was introduced in \cite{DBLP:journals/corr/HaDL16}, where the mechanism for dynamically modifying the weights of RNNs at each step was proposed. Similar to the discussion in \ref{subsec:static}, the primary purpose of the dynamic hypernetwork is originally not for conditioning, either. However, in our study, we apply this concept to VA modeling, dynamically adjusting the weights of RNNs based on control parameters, for potentially stronger conditioning.

In traditional RNNs, weights remain fixed throughout the sequence, meaning each step employs the same weights to generate results. In contrast, the DynamicHyper-RNN dynamically generates weights using another recurrent neural network, allowing for varying weights across each time step. We can use a smaller recurrent neural network, termed as \textbf{hyperRNN}, to generate the weights for the main recurrent neural network directly, denoted as \textbf{mainRNN}. 
%See Figure \ref{fig:dynamic} for an illustration. 
%We note that this approach may lead to memory-intensive operations and parameter explosion. In \cite{DBLP:journals/corr/HaDL16}, the authors proposed a weight scaling strategy instead of direct weight computation to mitigate these challenges. 
As shown in Figure \ref{fig:dynamic}, while StaticHyper only uses $\phi$ as input to generate the weights for RNN, DynamicHyper uses not only $\phi$ but also $h_{t-1}$ as the input to generate the weights. The input $x^{m}_{t}$ refers to the input audio signal, while the input $x^{p}_{t}$ is constructed by concatenating $h^{m}_{t-1}$ and the conditioning vector $\phi$, namely:
$$
x^{p}_{t} = \begin{pmatrix}
        h^{m}_{t-1} \\
        \phi
        \end{pmatrix}
$$

%The DynamicHyper-RNN includes 
The two components hyperRNN and mainRNN can be formulated with different equations. 
Using  the superscripts $p$ and $m$ to denote the variables for each of them, and omitting the bias term,
hyperRNN (``p'') and mainRNN (``m'') entail respectively:
\[ h^{p}_{t} = \tanh(W^{p}_{h}h^{p}_{t-1} + W^{p}_{x}x^{p}_{t}) \]
%where the superscript $p$ denotes the variables used for hyperRNN. 
%For mainRNN:
\[ h^{m}_{t} = \tanh(d_{h}(z_h) \odot W^{m}_{h}h^{m}_{t-1} + d_{z}(z_x) \odot  W^{m}_{x}x^{m}_{t}) \]
%where the superscript $m$ denotes the variables used for mainRNN, 
The functions $d_h$ and $d_z$ represent learnable transformations, and $z_h$ and $z_x$ are the features generated by the transformation from the hyperRNN. The $\odot$ operation means the element-wise multiplication. %The input $x^{m}_{t}$ refers to the input audio signal, while the input $x^{p}_{t}$ is constructed by concatenating $h^{m}_{t-1}$ and the conditioning vector $\phi$, namely:
%$$
%x^{p}_{t} = \begin{pmatrix}
%        # h^{m}_{t-1} \\
%        # \phi
%        # \end{pmatrix}
%$$
The features $z_h$ and $z_x$ resulting from the transformation of $h^{p}_{t}$ can be expressed as:
% \begin{alignat*}{2}
% \mathbf{f_{h}}(h^{p}_{t}) &= z_{h}, \quad && \mathbf{f_{x}}(h^{p}_{t}) &= z_{x}
% \end{alignat*}
\begin{align*}
\mathbf{f_{h}}(h^{p}_{t}) &= z_{h}, &
\mathbf{f_{x}}(h^{p}_{t}) &= z_{x}
\end{align*}
where the functions $\mathbf{f_{h}}$ and $\mathbf{f_{x}}$ can be implemented by neural network layers in practice. The hyperRNN  offers time-varying weights across each step, relaxing the share-weight strategy used in standard RNNs. We named our proposed models DynamicHyperRNN because the weights are dynamically modified at each step.

\subsection{Discussion}
From a machine learning point of view, all of the three proposed conditioning methods can actually be viewed as hypernetwork-based conditioning methods. For FiLM-based conditioning, the FiLM-ed generator serves as the hypernetwork, generating scaling and shifting parameters to interact with the feature map of the target network. For StaticHyper-based conditioning, the mechanism generates weights of RNNs using an MLP, with the MLP acting as the hypernetwork for the architecture. For DynamicHyper-based conditioning, smaller RNNs are employed to dynamically modulate weights based on control parameters, with these smaller RNNs serving as the hypernetwork.

\section{Experimental setup}
\subsection{Baseline}
%We establish the baseline by considering the modeling backbone and the conditioning mechanism to compare our proposed approach. We employ two types of modeling backbones: convolutional-based neural networks and recurrent-based neural networks. 

As our focus in this paper is to improve the conditioning mechanism for RNN-based VA modeling, the baseline approach to compare against our proposed methods would be the concatenation-based conditioning methods for RNNs.  However, besides RNNs, we are also interested in seeing how the combination of advanced conditioning methods with RNNs can rival CNN-based methods, focusing on the fidelity and audio quality of modeling instead of other aspects such as real-time factors and latency.

For CNN-based models, we adopt the micro-TCN \cite{steinmetz2022efficient} and GCN \cite{https://doi.org/10.48550/arxiv.2211.00497} as baselines, as these models have been utilized in previous studies. For RNN-based models, we pick LSTM and GRU, which have demonstrated impressive results in modeling distortion circuits 
%in prior research 
\cite{769f627fa4fe49569bd207f6b1d32dc3}. For TCN and GCN, we explore two conditioning mechanisms: the concatenation method and FiLM. 
%The former is widely employed in deep learning, while 
The latter has been utilized in \cite{steinmetz2022efficient} for modeling compressor control parameters. For RNN-based baselines, we only use the concatenation method for conditioning, the prevailing approach in the literature. 
In what follows, we will use the following naming principle: [\textbf{control-model}]. The former represents the control mechanism, and the latter represents the model used. For example, \textbf{FiLM-TCN} means the TCN backbone with FiLM conditioning.

\subsection{Model implementation details}
We implement all the models using PyTorch in this work. For fair comparison, we configure the hyperparameters of the implemented models in a way such that they share similar number of trainable parameters. Specifically, we set the hidden state size to 32 for the backbone GRU and LSTM models. The FiLM-GRU and FiLM-LSTM architectures use 2 layers of MLP with a hidden size of 32 as the FiLM-ed generator. Both StaticHyper-GRU and StaticHyper-LSTM utilize the MLP architecture for weight generation, featuring a hidden size of 8 and 3 layers. The DynamicHyper-GRU and DynamicHyper-LSTM employ smaller GRU and LSTM networks as the hypernetwork, each with a hidden size 8. The function that transforms the hidden state output by the smaller GRU or LSTM to the feature vector $z$, along with the transform function $d$ discussed in Section~\ref{dynamicchyper}, are implemented by 2 layers of MLP with a hidden size of 32. 

For TCN, we adopted the model architectures proposed in \cite{steinmetz2022efficient}. As for GCN, we followed the model architectures outlined in \cite{https://doi.org/10.48550/arxiv.2211.00497}. We chose a channel width of 24 for TCN and 32 for GCN so that they have similar number of parameters as the RNN-based models.
When modeling different devices, we varied the number of layers, kernel sizes, and dilation growth rates for the CNN-based models. This variation is because different receptive fields are suitable for modeling different types of effects. For instance, \cite{Damskgg2019RealTimeMO} suggests that a short receptive field can model distortion circuits, while \cite{https://doi.org/10.48550/arxiv.2211.00497} employs a long receptive field to model compressors.
To model the Boss OD-3, we stacked 10 layers with a kernel size of 3 and dilation growth of 2 as the hyperparameters, resulting in a receptive field of 2047 samples. For LA-2A, we stacked 9 layers with a kernel size of 5 and dilation growth of 3, leading to a receptive field of 39365 samples. For FiLM-TCN and FiLM-GCN, we utilized 3 layers MLP with a hidden size of 32 as the FiLM-ed generator. All the MLP layers have LeakyReLU activation functions with slope $0.1$ between each layer except for the last one.

\subsection{Training}

The implemented models are trained by minimizing carefully selected loss functions to achieve high-quality emulation. We employ a combination of time- and frequency-domain losses, including the L1 loss and the Multi-resolution STFT loss utilized in previous studies \cite{steinmetz2022efficient}. We use multi-resolution STFT loss, with three FFT window sizes: 128, 512, and 2048. All models are trained using the Adam optimizer with an initial learning rate of 1e-3, 100 epochs, and a batch size 32. The learning rate decays by half after five epochs of training. Lastly, we normalize the conditioning value to $-1$ to $1$ for all models.
For Boss OD-3, RNN-based models are trained using backpropagation through time every 2048 samples, reflecting the device's short-term memory characteristics. Conversely, for LA-2A, which concerns with long-range dependencies, the models are trained through every 8192 samples.
For CNN-based models, the input audio size is determined by ``the receptive field$+$buffer size$-$1,'' as outlined in \cite{steinmetz2022efficient}. Accordingly, we set the buffer size to 2048 samples and 8192 samples corresponding to the Boss OD-3 dataset and LA-2A dataset, respectively. 
\section{Evaluation}

\subsection{The proposed transient metric}

\begin{figure}[t] 
\includegraphics[width=1\columnwidth]{ 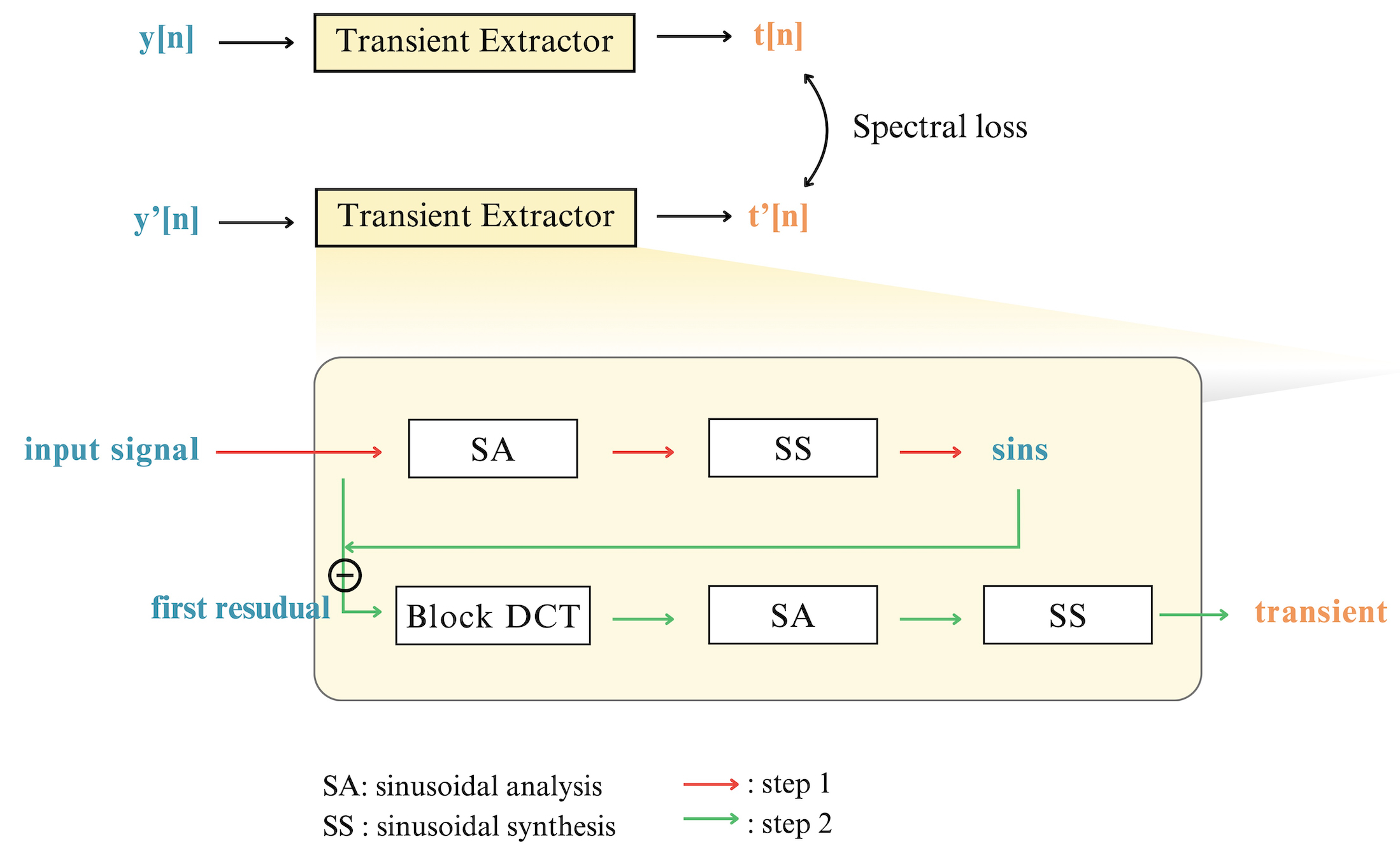} 
\caption{The diagram illustrates the proposed transient metric. The blue color represents the signal in the time domain, while the orange color signifies the signal in the discrete cosine transform (DCT) domain. The algorithm extracts the transient signal and calculates the spectral loss in the DCT domain.}
\label{fig:transient}
\end{figure}

To provide deeper insights into the performance of different models, we propose a novel metric to evaluate the transient reconstruction quality, inspired by the transient modeling synthesis method (TMS) outlined in \cite{verma2000extending}. 
While the TMS method has been there for more than two decades, to our best knowledge, the adaptation of TMS to construct a transient-centered objective metric for assessing the performance of VA modeling, or other audio generation tasks in general, has not been attempted before. Moreover, the TMS algorithm was not implemented in Python yet. We reimplement it based on sms-tools package\footnote{\url{https://github.com/MTG/sms-tools}} to facilitate its use in research today. As depicted in Figure \ref{fig:transient}, the TMS approach assumes that an audio signal can be decomposed into three components: sinusoids, transients, and noise. Here is a detailed breakdown: starting with the input audio signal, we initiate the process by applying sinusoidal analysis to retrieve the amplitude, frequency, and phase information to generate sinusoids. Subsequently, these sinusoids are subtracted from the original signal, resulting in the first residual signal, which includes the transient parts of the audio. This first residual signal is then transformed to the DCT domain, as it is easier to analyze transient signals in the DCT domain than in the time domain \cite{verma2000extending}. Next, we employ block-by-block sinusoidal analysis to synthesize the transient signal in the DCT domain, subsequently applying inverse DCT to restore it to the time domain. %We then subtract the time-domain transient signal from the first residual signal, yielding the second residual signal. Finally, we utilize a noise analysis method to isolate the final noise component \cite{604fb72e-5c74-33da-81dd-77994976bef2}. 

Following the TMS principle, we reconstruct the audio using TMS and isolate the DCT-domain transient part of the audio. Subsequently, we employ STFT loss to compute the transient's reconstruction error. We opt to utilize the DCT-domain transient part as input due to its ease of analysis via sinusoidal methods. 
\subsection{Objective evaluation}

\begin{table}[ht]
  \resizebox{\columnwidth}{!}{
  \begin{tabular}{llccccccr}
  \toprule
  \multirow{2}{*}{Model} & \multirow{2}{*}{Condition} & \multicolumn{6}{c}{OD-3 (Overdrive)} & \multirow{2}{*}{Params} \\
  \cmidrule(lr){3-8} 
  & & L1 & STFT & LUFS & CF & RMS & Transient & \\
  \midrule
  \multirow{5}{*}{LSTM} & Concat & 0.123 & 1.901 & 0.524 & 2.982 & 1.259 & 25.997 & 4769 \\
  & FiLM & 0.145 & 1.057 & 0.248 & 1.834 & 0.552 & 20.322 & 22561 \\
  & StaticHyper & 0.146 & 1.031 & 0.221 & 2.051 & 0.451 & 20.106 & 40449 \\
  & DynamicHyper & 0.149 & 0.695 & 0.199 & 2.431 & 0.402 & 12.968 & 21857 \\
  \midrule
  \multirow{5}{*}{GRU} & Concat & 0.120 & 1.933 & 0.455 & 2.932 & 1.096 & 27.338 & 3585 \\
  & FiLM & $\textbf{0.011}^{\ast}$ & $\textbf{0.536}^{\dag}$ & 0.176 & $\textbf{0.676}^{\ast}$ & 0.401 & 12.504 & 17217 \\
  & StaticHyper & 0.017 & 0.698 & 0.165 & 1.650 & $\textbf{0.318}^{\dag}$ & $\textbf{12.347}^{\dag}$ & 30369 \\
  & DynamicHyper & 0.150 & $\textbf{0.428}^{\ast}$ & $\textbf{0.075}^{\ast}$ & 0.883 & $\textbf{0.153}^{\ast}$ & $\textbf{11.308}^{\ast}$ & 20289 \\
  \midrule
  \multirow{3}{*}{TCN} & Concat & 0.033 & 0.928 & 0.305 & 1.177 & 0.671 & 27.634 & 21769 \\
  & FiLM & 0.044 & 0.698 & 0.338 & 0.894 & 0.842 & 33.678 & 29849 \\
  \midrule
  \multirow{3}{*}{GCN} & Concat & $\textbf{0.013}^{\dag}$ & 0.792 & 0.202 & $\textbf{0.776}^{\dag}$ & 0.447 & 19.103 & 19824 \\
  & FiLM & 0.149 & 0.672 & $\textbf{0.141}^{\dag}$ & 1.200 & 0.276 & 24.474 & 32368 \\
  \bottomrule
  \end{tabular}
  }
  \caption{Evaluation on the Boss OD-3 device test set. We denote the lowest error with $\ast$ and the second lowest error with $\dag$. Lower values indicate better quality for all metrics.}
  \label{tab:od3}
\end{table}

\begin{table}[ht]
  %\centering
  %\caption{Metrics}
  \label{tab:model_results_rotated}
  \resizebox{\columnwidth}{!}{
  \begin{tabular}{llccccccr}
  \toprule
  \multirow{2}{*}{Model} & \multirow{2}{*}{Condition} & \multicolumn{6}{c}{LA2A (Compressor)} & \multirow{2}{*}{Params} \\
  \cmidrule(lr){3-8} 
  & & L1 & STFT & LUFS & CF & RMS & Transient & \\
  \midrule
  \multirow{5}{*}{LSTM} & Concat & 0.105 & 1.326 & 1.328 & 2.471 & 3.182 & 26.315 & 4641 \\
  & FiLM & 0.105 & 0.630 & 1.446 & 2.359 & 3.119 & 21.592 & 22529 \\
  & StaticHyper & 0.012 & 0.633 & 1.468 & 2.046 & 3.347 & 22.438 & 40441 \\
  & DynamicHyper & 0.008 & 0.427 & 0.466 & 2.618 & 1.010 & 22.0662 & 21697 \\
  \midrule
  \multirow{5}{*}{GRU} & Concat & 0.108 & 0.507 & 0.716 & 2.081 & 1.640 & 21.002 & 3489 \\
  & FiLM & $\textbf{0.011}^{\dag}$ & 0.597 & 1.383 & $\textbf{2.006}^{\dag}$ & 3.081 & $\textbf{15.825}^{\ast}$ & 17185 \\
  & StaticHyper  & $\textbf{0.008}^{\ast}$ & $\textbf{0.371}^{\ast}$ & 0.543 & 2.386 & 1.211 & 20.437 & 30361 \\
  & DynamicHyper & 0.109 & $\textbf{0.377}^{\dag}$ & $\textbf{0.377}^{\dag}$ & $\textbf{1.919}^{\ast}$ & $\textbf{0.819}^{\dag}$ & $\textbf{19.826}^{\dag}$ & 20169 \\
  \midrule
  \multirow{3}{*}{TCN} & Concat & 0.099 & 0.579 & 0.743 & 2.223 & 1.499 & 31.485 & 28609 \\
  & FiLM & 0.036 & 0.447 & $\textbf{0.263}^{\ast}$ & 2.242 & $\textbf{0.585}^{\ast}$ & 30.827 & 35953 \\
  \midrule
  \multirow{3}{*}{GCN} & Concat & 0.102 & 0.657 & 1.416  & 2.278 & 3.079 & 20.745 & 25904\\
  & FiLM & 0.023 & 0.635 & 1.013 & 2.107 & 2.149 & 30.423 & 37408 \\
  \bottomrule
  \end{tabular}
  }
  \caption{Evaluation on the LA-2A device test set. We denote the lowest error with $\ast$ and the second lowest error with $\dag$. Lower values indicate better quality for all metrics.}
  \label{tab:la2a}
\end{table}

Besides transients, we employ commonly-adopted metrics to evaluate the model's performance from multiple perspectives. Regarding reconstruction quality, we employed the L1 loss and multi-resolution STFT loss. To assess loudness, we utilized the pyloudnorm \cite{steinmetz2021pyloudnorm} to estimate perceptual loudness error (LUFS). In measuring the system's dynamics, we utilized the crest factor (CF) and RMS error in the dB scale to estimate the dynamics error. 
%Finally, we employed the proposed transient metric to evaluate the reconstruction quality of the transient. 

\label{od3:info:descp}
Table \ref{tab:od3} presents the result of objective evaluation for the Boss OD-3, the nonlinear effect with short-term memory. Focusing on RNN-based models, all three proposed methods outperform the concatenation method in LSTM and GRU models across several metrics. Despite having fewer parameters, GRU-based models surpass LSTM-based models. DynamicHyper-GRU exhibits the best results across four metrics, showcasing the model's high capability. While FiLM-GRU and StaticHyper-GRU yield similar results, the former works better for frequency-related metrics, and the latter works better for loudness and dynamics metrics. Notably, all the proposed methods exhibit better transient reconstruction quality compared to the concatenation method. This suggests that our methods can enhance the model's ability to capture transients while the concatenation method struggles to handle them.

Table \ref{tab:od3} also displays the performance of CNN-based models. GCN outperforms TCN in terms of quality, and for both models, FiLM demonstrates a more effective conditioning ability than the concatenation method. Comparing the CNNs and RNNs, we see that the gating mechanism seems to be crucial for modeling overdrive effects. Upon comparing the performance between GRU and GCN, we observed that Concat-GRU yields worse quality than Concat-GCN and FiLM-GCN. However, our proposed models can achieve better or comparable results than FiLM-GCN. Regarding transient reconstruction, we observe that TCN and GCN struggle to model the transient, while the proposed conditioned methods with LSTM and GRU work better. This illustrates that the advanced conditioning mechanism can retain the advantages of RNNs (e.g., real-time usage) and improve model performance.
%and low latency

Table \ref{tab:la2a} presents the results of modeling the LA-2A, the nonlinear audio effect with time dependency. We focus on the result for RNN-based models first. Comparing GRU and LSTM, GRU demonstrates superior quality. Among the proposed three conditioning mechanisms, DynamicHyper-GRU consistently ranks as either the best or the second best, showcasing its strength. We conjecture that this is due to the behavior of time-varying weights, which resembles that of a compressor. Compressors can be interpreted as applying time-varying gain \cite{wright2022grey}. Another observation is that StaticHyper-GRU outperforms FiLM-GRU. As discussed in \ref{od3:info:descp}, FiLM-GRU is good at modeling frequency content. However, for compressors, dynamic information holds greater importance. From the perspective of model architecture, FiLM-GRU applies the same scaling and shifting coefficients to every model step, limiting its ability to handle time information effectively.

Table \ref{tab:la2a} also indicates that TCN outperforms GCN in modeling compressors. Comparing FiLM-TCN, StaticHyper-GRU, and DynamicHyper-GRU, we observe that FiLM-TCN achieves better results in loudness and dynamic metrics, while the others excel in STFT and transient performance. We infer that CNN-based models can capture longer-time information with a larger receptive field, which is beneficial for modeling compressors. However, Table \ref{tab:la2a} shows that it may struggle with handling transients.

\subsection{GFLOP analysis}
\label{sec:exp:compt}

\begin{table}[htbp]
\centering
\begin{tabular}{lc}
\hline
Models & GFLOPs \\
\hline
Concat-GRU & ~~0.325 \\
\hline
FiLM-GRU & ~~0.307 \\
StaticHyper-GRU & ~~0.003 \\
DynamicHyper-GRU & ~~1.907 \\
\hline
Concat-GCN & 59.388 \\ 
FiLM-GCN & 58.701 \\
\hline
\end{tabular}
\caption{The computational cost is measured in GFLOPs. We evaluate the processing of one-second audio samples at a sampling rate 48kHz for each model and calculate the GFLOPs. Smaller numbers indicate less  compute.}
\label{tab:flops}
\end{table}

%We analyzed the  compute of the proposed conditioning mechanisms, as our case study. 
To study the computational cost, we computed the floating point operations (FLOPs) for one-second 48kHz audio samples, using an open-source Python package.\footnote{\url{https://github.com/MrYxJ/calculate-flops.pytorch/tree/main}} We selected a conditioning vector size of 2, corresponding to the Boss OD-3 experiments in our work.  
As indicated in Table~\ref{tab:flops}, FiLM-GRU and StaticHyper-GRU demonstrate lower  compute than the concatenation method. This discrepancy arises because the concatenation method requires additional computation for the conditioning signal at each step. In contrast, with FiLM-GRU, the conditioning information remains fixed during inference, so the computation of scaling and shifting coefficients is done only once. However, we still perform element-wise multiplication at each step. Therefore, the  compute needed by FiLM-GRU and Concat-GRU is similar. In the case of StaticHyper-GRU, the pre-generated and fixed weights eliminate the need for further computation to handle the conditioning information, resulting in significantly lower compute than the previous two models.
Finally, DynamicHyper-GRU, despite demonstrating superior performance across several metrics, requires higher computational resources, approximately six times greater than Concat-GRU. This increased demand is due to the model's necessity to modulate weights over time.

\subsection{Spectrum analysis}

\begin{figure}[ht] 
\includegraphics[width=1\columnwidth]{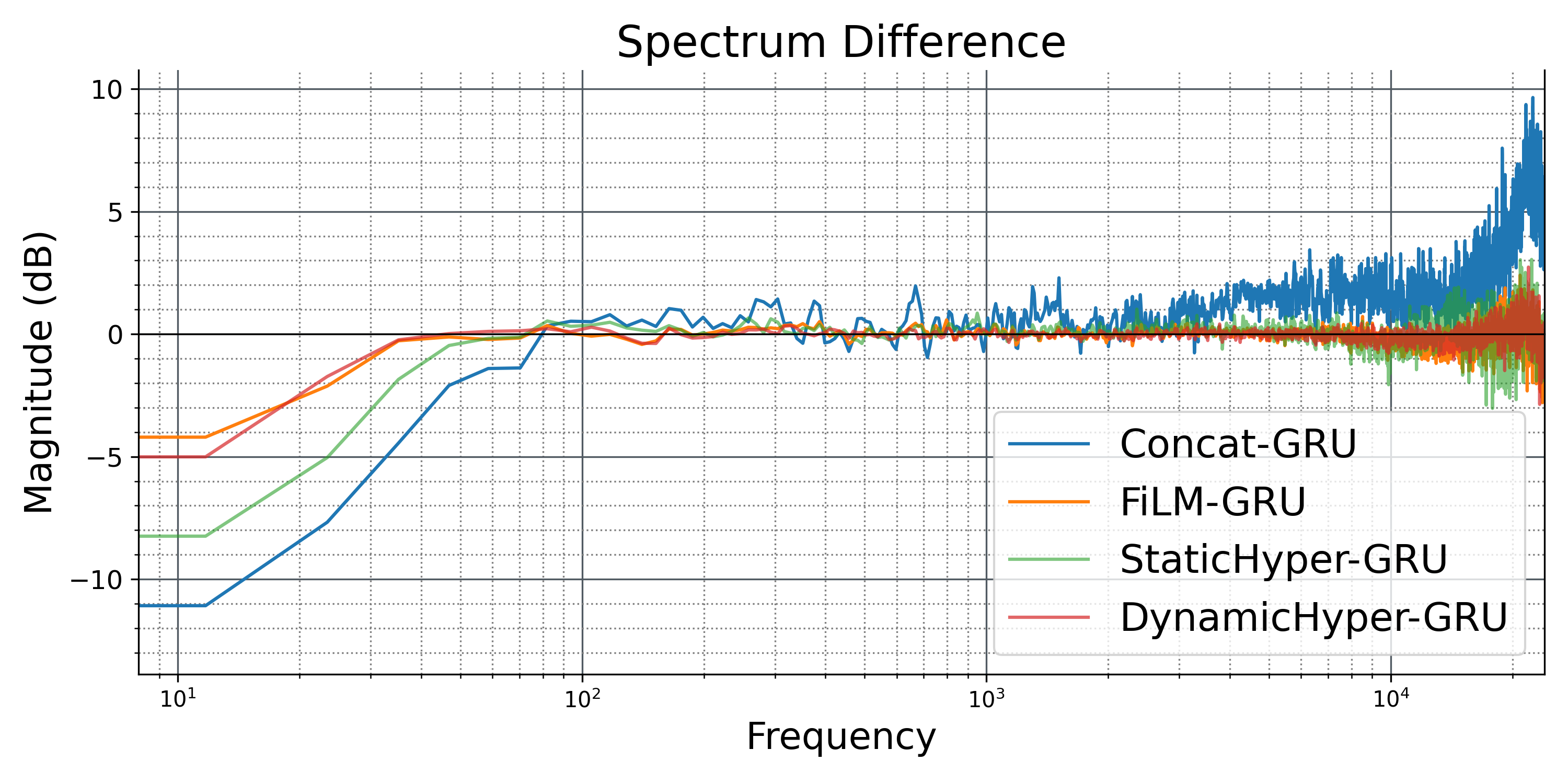} 
\caption{The diagram illustrates the spectrum difference observed in the Boss OD-3 test clips. All the proposed conditioning methods yield superior results compared to the Concatenation method.
}
\label{fig:spectrum}
\end{figure}

%We analyzed the result of the proposed models 
We analyzed the result of GRU with different conditioning methods 
%the GRU-based models 
in the frequency domain using clips from the Boss OD-3, with knob values setting to 3 and tone values 4. We consider only this knob setting here, because other cases lead to similar results. We computed the STFT loss for both the target and predicted clips and calculated the spectrum difference. As depicted in Figure \ref{fig:spectrum}, the concatenation method exhibits the greatest differences between the ground truth and the predictions, particularly in the high-frequency area.
In contrast, the proposed conditioning methods show fewer discrepancies. However, all methods have problems in accurately modeling high-frequency content. This limitation may stem from the aliasing effects or the neural network's capacity to handle high frequencies. Additionally, we observed significant discrepancies near the 0 frequency for all methods, indicative of DC bias errors.

\section{Discussion}

%In the upcoming section, we aim to provide some insights regarding our models. Through this discussion, we highlight potential areas for future improvement in our work and showcase 

We discuss below direction of future improvement, as well as how our model is empirically grounded and linked to deep learning (DL) and digital signal processing (DSP) principles.

%\subsection{FiLM and StaticHyper}
\textbf{FiLM and StaticHyper.}
To efficiently inject the conditioning information, we focus on the modulation potential, which is why DL can yield impressive results across various domains. At a higher level, our result suggests that FiLM and StaticHyper can be considered as lying on the two ends of the spectrum in terms of their modulation potential. FiLM employs fewer parameters than StaticHyper to modulate the intermediate feature map through linear transformation, limiting its modification ability to scale and shift coefficients. In contrast, StaticHyper allows a model to determine its weights directly based on conditioning information, fully leveraging the potential offered by such information.
Although FiLM uses fewer parameters, our compute analysis in Section \ref{sec:exp:compt} indicates that StaticHyper requires significantly fewer computational resources despite having more parameters. This phenomenon underscores the importance of analyzing the emulated models from multiple perspectives. A potential area for improvement lies in the conditioning representation. Our work normalized the conditioning value to $-1$ to 1 and fed it to FiLM or StaticHyper. Exploring alternative representations of the raw knob value may lead to better results. From a DL perspective, optimizing the conditioning representation can enhance the quality of results on unseen conditions.
%\subsection{DynamicHyper and time-varying}

\textbf{DynamicHyper and time-varying.}
We illustrate the advantage of DynamicHyper on two key factors. First, from a DL perspective, the model employs a relaxed weight-sharing strategy. This means the model can identify shared information across sequences while customizing weights for each step. Such a strategy enhances the expressivity of RNNs. Second, from a DSP viewpoint, the model offers a more intuitive representation of time-varying systems. While standard RNNs can produce different results at each step due to the different hidden states, this also limits the model's expressivity. DynamicHyper can greatly improve a model's expressivity by better exploiting time-dependent information. Moreover, we note that the way DynamicHyper is implemented in this paper is a straightforward case. There might be more advanced ways to utilize DynamicHyper, e.g., by using the signal from previous steps as the conditioning signal, that can be explored in future research.
%The advanced utilization of DynamicHyper remains unexplored. In our study, we provided the knob value as conditioning information to DynamicHyper, but it is also possible to use the signal from previous steps as the conditioning signal. We leave this exploration for future research.
%\subsection{Real-time implementation}

\textbf{Real-time implementation.} While we present in Section \ref{sec:exp:compt} a compute analysis, we do not analyze the real-time factors of our models yet, for it makes more sense if the models are implemented and optimzied in C++. We plan to do so in the future.

%While we provide the training compute analysis to our models, it is worth optimizing the model and implementing it in C++ in the future to further analyze the proposed models' real-time factors. 
%Although we did not implement our work in C++, 

\section{Conclusions}

This study has showcased advanced conditioning mechanisms for black-box virtual analog modeling, leveraging hypernetworks to enhance the modulation potential of neural networks. We assess our proposed methods across several dimensions, including recurrent units, device types, objective metrics, and compute analysis. In terms of recurrent units, we demonstrate their effective utilization with LSTM and GRU models. Regarding devices, our method surpasses the concatenation method across two types of nonlinear devices, namely those with short-term memory and time-dependent nonlinearity effects. We present results across several metrics, including time and frequency domain metrics, as well as a novel transient-related metric. Additionally, we calculate the FLOPs of the proposed methods, noting that FiLM-RNN and StaticHyper-RNN exhibit lower  computational cost. While DynamicHyper-RNN requires higher computational cost, it leads to better objective scores than the other models.

\section{Acknowledgments}
The authors would like to thank the support from the National Science and Technology Council of Taiwan (112-2222-E-002-005-MY2). We are grateful to Wei-Chieh Chou for helping with building the Boss OD-3 dataset with us. 

%\newpage
\nocite{*}
\bibliographystyle{IEEEbib}
\bibliography{DAFx24_tmpl} % requires file DAFx24_tmpl.bib

\begin{thebibliography}{10}

\bibitem{5280324}
David~T. Yeh, Jonathan~S. Abel, and Julius~O. Smith,
\newblock ``Automated physical modeling of nonlinear audio circuits for real-time audio effects,''
\newblock {\em IEEE Transactions on Audio, Speech, and Language Processing}, vol. 18, no. 4, pp. 728--737, 2010.

\bibitem{imort22ismir}
Johannes Imort, Giorgio Fabbro, Marco A.~Martinez Ramirez, Stefan Uhlich, Yuichiro Koyama, and Yuki Mitsufuji,
\newblock ``Distortion audio effects: Learning how to recover the clean signal,''
\newblock in {\em Proc. Int. Society for Music Information Retrieval Conf.}, 2022, pp. 218--225.

\bibitem{VolterraSeries10}
Thomas Hélie,
\newblock ``Volterra series and state transformation for real-time simulations of audio circuits including saturations: Application to the moog ladder filter,''
\newblock {\em IEEE Transactions on Audio, Speech, and Language Processing}, vol. 18, no. 4, pp. 747--759, 2010.

\bibitem{VolterraSeries18}
Simone Orcioni, Alessandro Terenzi, Stefania Cecchi, Francesco Piazza, and Alberto Carini,
\newblock ``Identification of volterra models of tube audio devices using multiple-variance method,''
\newblock {\em Journal of the Audio Engineering Society}, vol. 66, no. 10, pp. 823--838, 2018.

\bibitem{HMModel}
Felix Eichas and Udo Z{\"o}lzer,
\newblock ``Virtual analog modeling of guitar amplifiers with {W}iener-{H}ammerstein models,''
\newblock in {\em Proc. Annual Convention on Acoustics}, 2018.

\bibitem{steinmetz2020automatic}
Christian~J. Steinmetz, Jordi Pons, Santiago Pascual, and Joan Serrà,
\newblock ``Automatic multitrack mixing with a differentiable mixing console of neural audio effects,''
\newblock in {\em Proc. IEEE Int. Conf. Acoustics, Speech and Signal Processing}, 2021.

\bibitem{diffWhiteBox}
Fabián Esqueda, Boris Kuznetsov, and Julian~D. Parker,
\newblock ``Differentiable white-box virtual analog modeling,''
\newblock in {\em Proc. Int. Conf. Digital Audio Effects}, 2021, pp. 41--48.

\bibitem{Parker:2022.PhysicalModeling}
Julian~D. Parker, Sebastian~J. Schlecht, Rudolf Rabenstein, and Maximilian Schäfer,
\newblock ``Physical modeling using recurrent neural networks with fast convolutional layers,''
\newblock in {\em Proc. Int. Conf. Digital Audio Effects}, 2022.

\bibitem{eichas2018jaes}
Felix Eichas and Udo Zölzer,
\newblock ``Gray-box modeling of guitar amplifiers,''
\newblock {\em Journal of the Audio Engineering Society}, 2018.

\bibitem{wright2022grey}
Alec Wright and Vesa V{\"a}lim{\"a}ki,
\newblock ``Grey-box modelling of dynamic range compression,''
\newblock in {\em Proc. Int. Conf. Digital Audio Effects}, 2022, pp. 304--311.

\bibitem{colonel2022reverse}
Joseph~T. Colonel, Marco Comunità, and Joshua Reiss,
\newblock ``Reverse engineering memoryless distortion effects with differentiable waveshapers,''
\newblock {\em Journal of the Audio Engineering Society}, 2022.

\bibitem{miklanek2023neural}
Stepan Miklanek, Alec Wright, Vesa V{\"a}lim{\"a}ki, and Jiri Schimmel,
\newblock ``Neural grey-box guitar amplifier modelling with limited data,''
\newblock in {\em International Conference on Digital Audio Effects}. Aalborg University, 2023, pp. 151--158.

\bibitem{nercessian2021lightweight}
Shahan Nercessian, Andy Sarroff, and Kurt~James Werner,
\newblock ``Lightweight and interpretable neural modeling of an audio distortion effect using hyperconditioned differentiable biquads,''
\newblock {\em arXiv preprint arXiv:2103.08709}, 2021.

\bibitem{3e3037cffa094d3fb7a73dbc33ea0e9a}
Roope Kiiski, Fabian {Esqueda Flores}, and Vesa V{\"a}lim{\"a}ki,
\newblock ``Time-variant gray-box modeling of a phaser pedal,''
\newblock in {\em Proc. Int. Conf. Digital Audio Effects}, 2016.

\bibitem{app10020638}
Marco~A. Mart{\'\i}nez~Ram{\'\i}rez, Emmanouil Benetos, and Joshua~D. Reiss,
\newblock ``Deep learning for black-box modeling of audio effects,''
\newblock {\em Applied Sciences}, vol. 10, no. 2, pp. 638, 2020.

\bibitem{Damskgg2019RealTimeMO}
Eero-Pekka Damsk{\"a}gg, Lauri Juvela, and Vesa V{\"a}lim{\"a}ki,
\newblock ``Real-time modeling of audio distortion circuits with deep learning,''
\newblock in {\em Proc. Sound and Music Computing Conf.}, 2019.

\bibitem{8682805}
Eero-Pekka Damskägg, Lauri Juvela, Etienne Thuillier, and Vesa Välimäki,
\newblock ``Deep learning for tube amplifier emulation,''
\newblock in {\em Proc. IEEE Int. Conf. Acoustics, Speech and Signal Processing}, 2019, pp. 471--475.

\bibitem{8683529}
Marco~A. Martínez~Ramírez and Joshua~D. Reiss,
\newblock ``Modeling nonlinear audio effects with end-to-end deep neural networks,''
\newblock in {\em Proc. IEEE Int. Conf. Acoustics, Speech and Signal Processing}, 2019.

\bibitem{steinmetz2022efficient}
Christian~J. Steinmetz and Joshua~D. Reiss,
\newblock ``Efficient neural networks for real-time modeling of analog dynamic range compression,''
\newblock {\em Journal of the Audio Engineering Society}, 2022.

\bibitem{https://doi.org/10.48550/arxiv.2211.00497}
Marco Comunità, Christian~J. Steinmetz, Huy Phan, and Joshua~D. Reiss,
\newblock ``Modelling black-box audio effects with time-varying feature modulation,''
\newblock {\em arXiv preprint arXiv:2211.00497}, 2022.

\bibitem{6567472}
John Covert and David~L. Livingston,
\newblock ``A vacuum-tube guitar amplifier model using a recurrent neural network,''
\newblock in {\em Proc. IEEE Southeastcon}, 2013.

\bibitem{schmitz2018real}
Thomas Schmitz and Jean-Jacques Embrechts,
\newblock ``Real time emulation of parametric guitar tube amplifier with long short term memory neural network,''
\newblock {\em arXiv preprint arXiv:1804.07145}, 2018.

\bibitem{Zhang2018AVG}
Zhichen Zhang, Edward Olbrych, Joseph Bruchalski, Thomas~J. McCormick, and David~L. Livingston,
\newblock ``A vacuum-tube guitar amplifier model using long/short-term memory networks,''
\newblock {\em SoutheastCon}, 2018.

\bibitem{769f627fa4fe49569bd207f6b1d32dc3}
Alec Wright, Eero-Pekka Damsk{\"a}gg, and Vesa V{\"a}lim{\"a}ki,
\newblock ``Real-time black-box modelling with recurrent neural networks,''
\newblock in {\em Proc. Int. Conf. Digital Audio Effects}, 2019.

\bibitem{NeuralODE}
Jan Wilczek, Alec Wright, Vesa V{\"a}lim{\"a}ki, and Emanu{\"e}l Habets,
\newblock ``Virtual analog modeling of distortion circuits using neural ordinary differential equations,''
\newblock {\em arXiv preprint arXiv:2205.01897}, 2022.

\bibitem{rethage2018wavenet}
Dario Rethage, Jordi Pons, and Xavier Serra,
\newblock ``A {WaveNet} for speech denoising,''
\newblock in {\em Proc. IEEE Int. Conf. Acoustics, Speech and Signal Processing}, 2018.

\bibitem{perez2017film}
Ethan Perez, Florian Strub, Harm de~Vries, Vincent Dumoulin, and Aaron Courville,
\newblock ``{FiLM}: Visual reasoning with a general conditioning layer,''
\newblock {\em arXiv preprint arXiv:1709.07871}, 2017.

\bibitem{2021hyeprconv}
Alexander Richard, Dejan Markovic, Israel~D. Gebru, Steven Krenn, Gladstone~Alexander Butler, Fernando Torre, and Yaser Sheikh,
\newblock ``Neural synthesis of binaural speech from mono audio,''
\newblock in {\em Proc. Int. Conf. Learning Representations}, 2021.

\bibitem{DBLP:journals/corr/HaDL16}
David Ha, Andrew~M. Dai, and Quoc~V. Le,
\newblock ``Hypernetworks,''
\newblock {\em arXiv preprint arXiv:1609.09106}, 2016.

\bibitem{simionato2022deep}
Riccardo Simionato and Stefano Fasciani,
\newblock ``Deep learning conditioned modeling of optical compression,''
\newblock in {\em Proc. Int. Conf. Digital Audio Effects}. DAFx Board, 2022.

\bibitem{verma2000extending}
Tony~S Verma and Teresa~HY Meng,
\newblock ``Extending spectral modeling synthesis with transient modeling synthesis,''
\newblock {\em Computer Music Journal}, pp. 47--59, 2000.

\bibitem{oord2016wavenet}
Aaron van~den Oord, Sander Dieleman, Heiga Zen, Karen Simonyan, Oriol Vinyals, Alex Graves, Nal Kalchbrenner, Andrew Senior, and Koray Kavukcuoglu,
\newblock ``{WaveNet}: A generative model for raw audio,''
\newblock {\em arXiv preprint arXiv:1609.03499}, 2016.

\bibitem{simionato2023fully}
Riccardo Simionato and Stefano Fasciani,
\newblock ``Fully conditioned and low-latency black-box modeling of analog compression,''
\newblock in {\em Proc. Int. Conf. Digital Audio Effects}. DAFx Board, 2023.

\bibitem{pedroza2022egfxset}
Hegel Pedroza, Gerardo Meza, and Iran~R Roman,
\newblock ``{EGFxSet}: Electric guitar tones processed through real effects of distortion, modulation, delay and reverb,''
\newblock {\em ISMIR Late Breaking Demo}, 2022.

\bibitem{stein2010automatic}
Michael Stein, Jakob Abe{\ss}er, Christian Dittmar, and Gerald Schuller,
\newblock ``Automatic detection of audio effects in guitar and bass recordings,''
\newblock in {\em Audio Engineering Society Convention 128}. Audio Engineering Society, 2010.

\bibitem{hawley2019signaltrain}
Scott~H. Hawley, Benjamin Colburn, and Stylianos~I. Mimilakis,
\newblock ``Signaltrain: Profiling audio compressors with deep neural networks,'' 2019.

\bibitem{martinez2020deep}
Marco~A Mart'{i}nez~Ram'{i}rez, Emmanouil Benetos, and Joshua~D Reiss,
\newblock ``Deep learning for black-box modeling of audio effects,''
\newblock {\em Applied Sciences}, vol. 10, no. 2, pp. 638, 2020.

\bibitem{Sherstinsky_2020}
Alex Sherstinsky,
\newblock ``Fundamentals of recurrent neural network ({RNN}) and long short-term memory ({LSTM}) network,''
\newblock {\em Physica D: Nonlinear Phenomena}, vol. 404, 2020.

\bibitem{steinmetz2021pyloudnorm}
Christian~J. Steinmetz and Joshua Reiss,
\newblock ``pyloudnorm: {A} simple yet flexible loudness meter in {P}ython,''
\newblock {\em Journal of the Audio Engineering Society}, 2021.

\end{thebibliography}

%\section{Appendix: Margin Check}

\end{document}